\documentclass[prl,final,twocolumn,superscriptaddress]{revtex4}

\usepackage{graphicx}

\def\tmp{$(TMTSF)_{2}PF_{6}$}\,
\def\tmc{$(TMTSF)_{2}ClO_{4}$}\,
\def\tmx{$(TMTSF)_{2}(ClO_{4})_{(1-x)}(ReO_{4})_{x}$}\,
\def\tms{$(TMTSF)_{2}(AsF_{6})_{(1-x)}(SbF_{6})_{x}$}\,
\,
\def\R{$ReO_{4}^{-}$}

\begin{document}

\title{Suppression of  superconductivity by non-magnetic disorder in 
 the organic superconductor \tmx}
\author{ {N.Joo$^{1,2}$, P.Auban-Senzier$^{1}$, C.R.Pasquier$^{1}$, P.Monod$^{3}$, D.J\'{e}rome$^{1}$ and K.Bechgaard$^{4}$}\\
$^{1}$\textit{Laboratoire de Physique des Solides (CNRS,UMR 8502), Universit\'e  Paris-Sud, 91405 Orsay, France}\\ 
$^{2}$\textit{Facult\'e des Sciences de Tunis, LPMC, Campus Universitaire , 1060 Tunis, Tunisie}\\
$^{3}$\textit{Laboratoire de Physique des Solides (CNRS,UPR 5 ), ESPCI, 75231 Paris-Cedex 05, France}\\ 
$^{4}$\textit{Polymer Department, Ris\o \, National Laboratory, 4000 Roskilde, Denmark}}

\begin{abstract}
\vspace{0.3cm}
We present a study of the superconducting properties $(T_c \, $and $H_{c2})$ in the solid solution \tmx\, with a \R nominal
concentration up to
$x=6\%$. The dramatic suppression of $T_c$ when the residual resistivity is increased upon alloying with no modification of the
Fermi surface is the signature of non-conventional superconductivity . This behaviour strongly supports \textit{p} or \textit{d}
wave pairing in  quasi one dimensional organic superconductors. The determination of the electron lifetime in the normal state at
low temperature confirms that a single particle Drude model is unable to explain the temperature dependence of the conductivity and
that a very narrow zero frequency mode must be taken into account for the interpretation of the transport properties.

\end{abstract}


\maketitle

Organic superconductivity has been around for nearly 25 years since its
discovery in \tmp \cite{Jerome80}. However very little is still known
experimentally regarding the symmetry of the superconducting pairing.
An early theoretical work of Abrikosov \cite{Abrikosov83}  performed
within the self consistent approximation has concluded that both
scenarios could be envisaged for quasi one dimensional (Q-1D) conductors namely, singlet or triplet
spin pairing depending on the properties of the electron-electron interaction. It was also anticipated 
that the stability of the triplet state would be very sensitive to the presence of non
magnetic impurities.

An other approach based on the exchange of antiferromagnetic fluctuations
between neighbouring chains has led to an exotic  $d$-like pairing
\cite{Bourbonnais88}. This model is singlet in the spin sector and leads to a superconducting gap which exhibits a sign
reversal along the open Fermi surface of these Q-1D conductors. Experiments available up today
are unable  to settle without ambiguity the question of  the pairing symmetry in organic conductors. As far as one dimensional superconductors
are concerned, some support in favour of triplet pairing was claimed from the analysis of critical fields in \tmc \,and \tmp \, in terms of
pure type II superconductivity \cite{Gorkov85} and also from the observation of a divergence of the upper critical field  at low temperature
for fields parallel to the intermediate direction b$^{\prime}$ with
$H_{c2}(T)$ exceeding greatly the paramagnetic (Pauli) limit value in \tmp\cite{Lee97}. In addition, the spin
susceptibility found to be temperature independent in the superconducting phase and equal to its normal state
value for \tmp \,has been taken as an other evidence in favour of triplet pairing\cite{Lee02}.
However, both experiments may not be  fully convincing  since in the former the critical field has been
investigated in a pressure domain close to the low pressure spin density wave instability where
superconductivity is inhomogenous\cite{Lee03,Vuletic02} whereas for the latter it is unclear whether the
NMR signal needed for the Knight shift measurement refers to superconducting or to normal electrons in 
this superconducting state exhibiting a vortex structure.
For \tmc , another member of the Bechgaard salts in which superconductivity is stable under ambient
pressure
\cite {Bechgaard81} the situation remains also unclear. The finding of a power law dependence of the
relaxation rate in the superconducting phase down to $T_{c}/2$ led Takigawa \textit{\textit{et-al}}\cite{Takigawa87}
to suggest the existence of lines of zeros for the superconducting gap on the Fermi surface\cite{Hasegawa87} while
thermal conductivity data down to $T_{c}/7$ provided compelling evidences for a nodeless gap
\cite{Belin97}. As far as the 2-D organic superconductors built on the $BEDT-TTF$ molecule are concerned  the actual symmetry of the
superconducting wave function is still controversial although far more experimental  studies have been conducted. Most NMR investigations
support a spin-singlet pairing\cite{Mayaffre95, deSoto95, Kanoda96}. Recent specific heat measurements performed in two members of the
$BEDT-TTF$ family have shown a fully gapped order parameter\cite{Elsinger00,Muller02}  \textit{at variance} with  early heat conduction
\cite{Belin98} and penetration depth measurements\cite{Pinteric00} providing evidences for nodes in the gap. This contradictory situation might be reasonably
resolved by recent penetration depth  experiments pointing out the critical influence of the ethylene groups ordering on the existence of low energy
electron excitations\cite{Pinteric02}. We may notice that $T_{c}$ in these 2-D superconductors is very sensitive to the existence of intrinsically non-magnetic
disorder\cite{Powell04}.

The quantum states of the partners entering into the formation of a Cooper pair of a BCS \textit{s}-wave
superconductor are related to each other by a time reversal symmetry. Hence no pair breaking is
expected from the scattering of electrons against spinless impurities \cite{Anderson59}. Experimentally,
this property has been verified in non-magnetic dilute alloys of \textit{s}-wave superconductors and
brought a strong support for the BCS model of conventional \textit{s}-wave superconductors. However,
the condition for time reversal symmetry is no longer met for the case of \textit{p}-wave pairing and
consequently $T_c$ for these superconductors should be strongly affected by even a small amount of
such non-magnetic scattering. It is the  extreme dependence of the critical temperature of
$Sr_{2}RuO_4$
\cite{Mackenzie98} on non-magnetic disorder which has provided a  strong support in favour of triplet
superconductivity in this compound. 
It is also the  remarkable sensisitivity  of organic
superconductivity to irradiation\cite{Bouffard82,Choi82} which led Abrikosov to suggest the possibility of
triplet pairing in these materials\cite{Abrikosov83}.
Although irradiation was recognized to be an excellent method for the introduction of defects in a
controlled way\cite{Zuppirolli87}, defects thus created can be magnetic\cite{Sanquer85} and
the suppression of superconductivity by irradiation induced defects  as a signature of
non-conventional pairing must be taken with "a grain of salt"  since local moments can also  act  as
strong pair-breakers on \textit{s}-wave superconductors.

Using chemistry to break the crystalline invariance is an other way for the creation of local
non-magnetic scatterers.  This has been achieved by the substitution of $TMTSF$ for $TMTTF$ molecules
in $TM_{2}X$ salts with $X=PF_6$ \cite{Coulon82} and $X=ClO_4$\cite{Johannsen85}. However in both
situations cationic alloying involves drastic modification of the normal state electronic properties
since the SDW transition of \tmp \,is quickly broadened and pushed towards higher temperature upon
alloying\cite{Mortensen84}. Consequently, such a cationic alloying may not be  the best case to test  the robustness of superconductivity
against an enhanced scattering. A softer way of introducing disorder on the cation stack has been developed with
the series (trimethyl-$TSF_{(1-x)} TMTSF_{x})_{2}X$ where a drastic suppression of
superconductivity of $X=ClO_4$ is observed with 10\% alloying while the SDW instability of $X=PF_6$
which involves an order of magnitude higher in energy is left unperturbed\cite{Johannsen85}. 

Leaving the cation
stack uniform, scattering centers can also be created on the anion stacks. The role of the anion
stack on the ground state is enhanced as soon as the anion which is located at an inversion center of the
structure  does not possess a central symmetry\cite{Pouget87}. This is the case in particular for tetrahedral anions
such as $X=ClO_4$ which order  at low temperature ($T_{AO}$=24K) in line with 
entropy minimization. As anion reorientation requires a  tunneling process between two states at equal energy but
separated by  a large potential barrier, the dynamics of orientation is a slow process at low
temperature. Hence, for  samples slowly cooled through $T_{AO}$ (in the so-called  R-state) the orientation of the anions is
uniform along the stacking axis  but alternate along the  b-direction leading in-turn to a doubled
periodicity with a concomitant opening of an energy gap on the Fermi surface at  $\pm \pi /2b$ and the creation of two
sheets of open Fermi surfaces at +k$_F$ and -k$_F$ respectively. Fast cooled samples reach low
temperature in a state (the Q-state) where orientational  disorder is frozen-in (\textit{i.e.} a mixture of both anion
orientations) and superconductivity is depressed\cite{Tomic83a,Garoche82}. When the cooling rate is fast enough 
(quenched state) full disorder is preserved  and the single-sheet Fermi surface of the high temperature
sructure prevails at low temperature leading in turn to the instability of the metallic phase against a SDW ground
state at $T_{SDW}$=5K\cite{Tomic83a}. Furthermore, it has been shown that neither the Pauli susceptibility\cite{Tomic83a,Tomic86} nor the density of
states\cite{Garoche82} of the normal phase are affected by the orientational disorder introduced by the fast cooling procedure as long as a superconducting
ground state is observed.  An other  approach for the introduction of anionic disorder has been successful with the synthesis of an anionic solid solution
involving anions of similar symmetry. For centrosymmetrical anions a suppression of the SDW state has been observed in
\tms
\cite{Traetteberg93} but the effect on superconductivity which would have required a high pressure has
not been studied in details. As far as the non-centrosymmetrical anions are concerned,   the early studies by
Tomi\'c \textit{et-al}
\cite{Tomic83b}in\tmx \, have shown that both the low temperature conductivity and the  transition towards
superconductivity are very strongly affected by alloying although X-ray investigations have revealed
that long range order is preserved up to 3\% $ReO_{4}^{-}$ with a correlation length $\xi$$_a$  $>$ 200\AA
\cite{Ravy86}.

In the present paper we have investigated the influence of non-magnetic disorder on the
superconducting properties of the solid solution \tmx \, with different cooling rates down to 100mK.
>From the measurement of $T_c$ and of the upper critical field at different $x$ and  cooling
rates we have derived the relation between the coherence length and the critical temperature which is
\textit{at variance} with the expectation in a conventional dirty type II superconductor.

\vspace{2mm}
Single crystals of  the solid solution \tmx \, with $x$ in the range $0\leq x\leq 0.06$ have been
prepared with the usual electrocrystallization technique. As
the relative concentration of $ReO_{4}^{-}$ and $ClO_{4}^{-}$ anions in the crystals could differ from the nominal
concentration of the solution a microprobe analysis has been attempted. However the sensitivity of such a  technique with respect to the
$ReO_{4}^{-}$ concentration is rather limited\cite{note1}.  Only the $x=0.06$ sample gave a rhenium peak with a  signal to noise ratio which
was large  enough to enable a calibration. We found that the actual concentration is about 30\% smaller than the value expected from the
nominal concentration with a large error bar. We shall see in the following that the nominal concentration is not so relevant for the purpose
of our experiment.

\begin{figure}[htbp]
\centerline{\includegraphics[width=0.9\hsize]{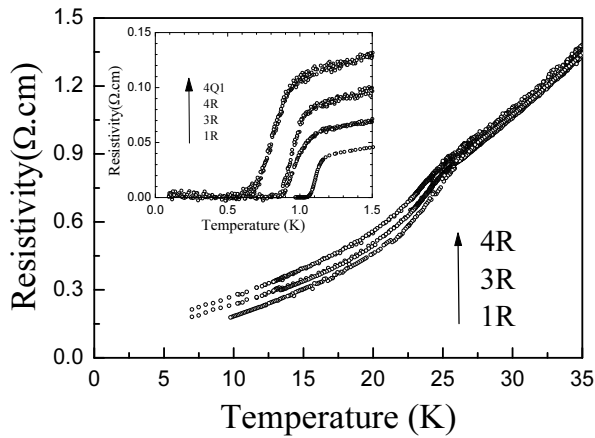}}
\caption{Temperature dependence of the resistivity in the low temperature domain for three typical \tmx samples with different
$ReO_{4}^{-}$ concentrations in the relaxed  state. The transition towards superconductivity is  displayed in the insert for 
non-alloyed (1) and alloyed (3 and 4) samples in relaxed (R) or quenched (Q) states (Q1 corresponds to the cooling rate of $-4K/mn)$.}
\label{fig-rhovsT}
\end{figure}
\begin{figure}[htbp]
\centerline{\includegraphics[width=0.9\hsize]{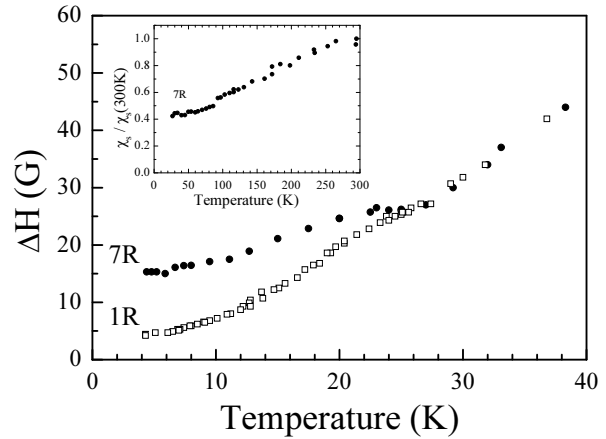}}
\caption{ESR linewidth of pure (1R) and doped (7R) samples. The difference in temperature dependences of the two samples below $T_{AO}$ is due to the
enhancement of  the electronic scattering rate in the doped system. The anion ordering is somewhat smeared by a small temperature lag ($\approx$ 2 K)
existing between the thermometer and the sample in the range 24K-12K. The inset shows the ESR susceptibility of the doped sample normalized at 300K. The
temperature dependence is in close agreement with that of \tmc published in the literature. }
\label{figure2DH.eps}
\end{figure}

Conductivity measurements have been conducted using a low frequency lock-in detection with the
current flowing along the $c^\star$ axis of the crystals. This direction has been chosen for transport
studies because it allows a reliable cooling without visible cracks and also because the $a-b^\prime$ plane 
due to the high transport
anisotropy provides a well defined equipotential surface for the resistance measurements . Contacts were
made with silver paste on two gold pads evaporated on each parallel side of  crystals with typical sizes  $1\times 0.3\times 0.15 mm^3$.
The low temperature was provided by a dilution refrigerator designed for convenient operation between 100mK
and 30K with a sample holder able to measure two samples simultaneously. For the comparison of the low temperature resistivity between different samples
which is needed in this work we did not rely on their geometrical factors relating resistance to resistivity. This factor is always difficult to determine with
much accuracy due to uncertainties in the measurement of the sample dimensions.  Instead, we chose
$28
\Omega  cm$  for the value of the room temperature transverse resistivity $\rho_{c^\star}$,  an average derived from the study of 15 different samples.
Subsequently we have normalized all  measurements of resistance \textit{versus} temperature   against this average resistivity. This procedure is justified by
the application of the Mathiessen's law which is particularly relevant in this experiment since the resistance  ratio between ambient and helium temperature 
even in the most impure samples was at least 100.  The ESR linewidth data displayed in fig. (\ref{figure2DH.eps}) bring a further justification  for the
procedure used to compare resistivity between  different samples.  Furthemore, we chose to characterize the electronic life time  at low temperature by the
residual resistivity
$\rho_0$ which we obtain as  the constant parameter in a polynomial fit of the  temperature dependence
of the  resistivity in the normal state (instead of the nominal concentration of the solid solution). The
resistivity is linear (quadratic) in temperature below (above) 10K \cite{noteT2}.

Reliable data require a good control of the anion ordering, therefore the procedure for low temperature
measurements was the following: the sample was first slowly  cooled down to 4.2K $(\leq 0.2K/mn)$, then 
warmed up to 30K and subsequently cooled again to low temperature with an adequate  control of the cooling rate
through the anion ordering temperature. 
The cooling rate used to stabilize the R-state of \tmx in our experiments is -0.05K/mn(3K/h).This rate is indeed
much lower than the lowest rate used earlier to study the R-state (12 and 5K/h)\cite{Tomic83b}.
Figure (\ref{fig-rhovsT}) displays  resistivity data in the low temperature domain. In the present study we have
focused our attention on the samples with $0\leq x\leq 0.06$ measured with  cooling  rates  slow enough
in order to preserve a metallic behaviour  ($\delta\rho / \delta T>0$) over the entire temperature regime.
According to fig (\ref{fig-rhovsT}) the signature of anion ordering is visible at $T_{AO}$= 24K and 
survives although  smeared out upon alloying up to the nominal concentration of $x=0.06$.  No  decrease of $T_{AO}$ larger than 1K or so has been
noticed in agreement with X-ray investigations\cite{Ravy86,Ilakovac97,note2}. In
addition we notice that the residual resistivity increases  with the concentration of \R defects in the R-phase  and also with the cooling rate for a given
sample in agreement with Mathiessen's law, \textit{see} insert in fig.(\ref{fig-rhovsT}). We also noticed that much lower cooling rates are necessary to reach
a R state in highly defective samples in agreement with the increase of the annealing time requested by annealing close to $T_{AO}$ as measured by Tomi\'c
\cite{Tomic83a}. 

Electron paramagnetic resonance studies have been carried on in a X-band spectrometer equipped with a variable temperature cryostat able to control the cooling
rate down to 4.2 K. Typical results are shown on fig (\ref{figure2DH.eps}) where the peak to peak linewidths ($\Delta$H) of pure (1R) and $x=0.06$ (7R) samples
are displayed against temperature. Figure (\ref{figure2DH.eps}) reveals a temperature dependence as expected for the resonance of a conduction electron spin
resonance (CESR) in metals, namely  $\Delta$H $\propto$ electronic scattering rate (resistivity)\cite{Beuneu78}. The magnetic field is along the $b$ axis of
the single crystals and the g factor has been found temperature independent. The CESR data show that Mathiessen's law applies fairly well to  $\Delta$H  since
similar increases of linewidth and resistivity are observed upon doping. Furthermore, $ReO_4$ doping has no effect on the integrated electron susceptibility
temperature dependence as it should if the doped samples are free from magnetic impurities, see inset.

For all samples the superconducting transition can be detected by a drop of the resistance to zero.
 The onset temperature was defined by  the intercept between the  high
temperature extrapolation and the linear fit in the transition region. In the pure sample it amounts to 1.18K with a  (10\%-90\%) width of
0.1K.

\begin{figure}[htbp]
\centerline{\includegraphics[width=0.9\hsize]{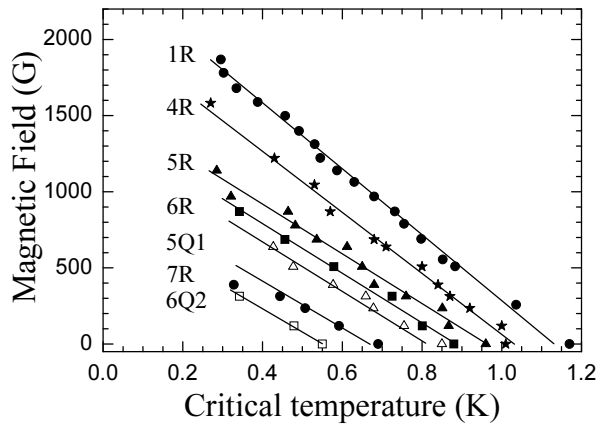}}
\caption{Critical temperature \textit{vs} magnetic field  for five crystals of \tmx with different $T_c$, Q1 and
Q2 refer to quenched samples with cooling rates of $-4K/mn$ and $-26K/mn$ respectively.}
\label{Hc2vsT}
\end{figure}

When a magnetic field is applied along the $c^{\star}$ direction, the whole $R(T)$ curve is pushed towards low
temperature. Hence $H_{c2}(T)$ can be derived for different samples and different cooling rates, see fig
(\ref{Hc2vsT}). According to the data of fig (\ref{Hc2vsT}) the temperature dependence of $H_{c2}(T)$ is linear
close to $T_c$ as expected from the Ginzburg-Landau theory \cite{degennes66} and a value for $T_c$ can thus be determined 
 by the zero field extrapolation of the critical field\cite{note3}.

\begin{figure}[htbp]
\centerline{\includegraphics[width=0.9\hsize]{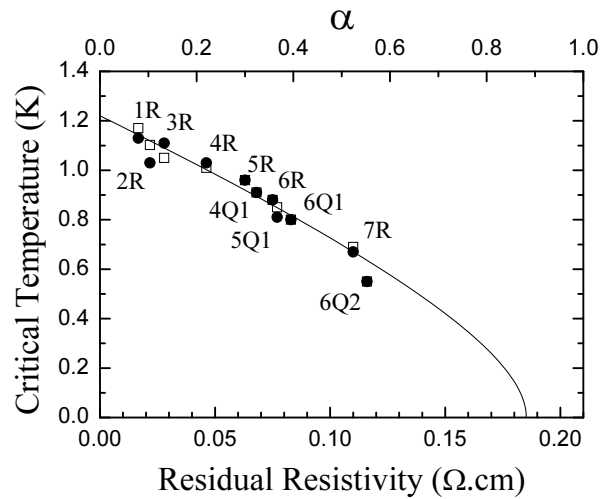}}
\caption{Superconducting critical temperature  derived from the zero field intercept of $H_{c2}(T)$ in fig (\ref{Hc2vsT}) (dots) and also from the zero field
resistivity data (squares)  as a function of the residual resistivity for samples with different amounts of disorder either chemical or(and) orientational.  The
solid line is a least square fit of the digamma pair-breaking function to the data which reaches zero at $\alpha = 0.88$.}
\label{Tcvsrho}
\end{figure}

The central result of this work is shown on fig (\ref{Tcvsrho}) where $T_c$ is plotted against the residual resistivity in
the transverse direction. On this figure the data  which are displayed come from different samples in the R state and also
 from quenched samples  preserving the metallic behaviour and superconductivity at
low temperature. It can be  inferred that the residual resistivity is proportional to the inverse electron
life time $1/\tau_{}$ (or inverse mean free path $1/l$). This is the case for coherent transport along $c^{\star}$
\cite{Korin88} but also for incoherent transport
since then we also have $\rho_{0c\star} \propto 1/\tau_{}$ \cite{Weger78,Friedel82}. The relation between the transverse resistivity and the in-chain inverse
electron lifetime is still valid in case of incoherent transverse transport as shown recently by the perturbative derivation of the transverse conductivity in
the tunnelling approximation\cite{Moser98,Georges00}. On account of the small value of the overlap along the $c$ direction for \tmc , the $c$-axis transport is
probably at the border between coherence and incoherence. However, the observation of a narrow angular magnetoresistance peak when a large magnetic field is
applied perpendicular to the $c$ axis supports the possibility of coherent transport at low temperature\cite{Moses99}.  Figure (\ref{Tcvsrho}) shows that the
extra scattering due to orientational disorder in fast cooled samples adds to the residual scattering due to the chemical defects. From the signature of the
anion ordering at
$T_{AO}$ we can be confident that the density of states at the Fermi energy is not changed in the solid solution. In addition  an effect of volume change can
also  be ruled out to explain the data on fig (\ref{Tcvsrho}) since the substitution of $ReO_{4}^{-}$ to $ClO_{4}^{-}$ should expand the
unit cell and in turn increase $T_c$ \textit{at variance} with the data\cite{Ferraro87}. Consequently, the suppression of $T_c$
must be related to the enhancement of the scattering rate in the solid solution. 
The additional scattering cannot be ascribed to magnetic scattering according to our  ESR studies. Thus, our
results cannot be reconciled with the picture of a superconducting gap keeping a constant sign over the whole
$(\pm k_F)$ Fermi surface and require a picture of pair breaking in a superconductor with an unconventional gap
symmetry. The conventional pair breaking theory for magnetic impurities in usual superconductors can thus
be generalized to the case of non-magnetic impurities in unconventional materials and $T_c$ reads\cite{Maki04,Larkin65},
$$ln(\frac{T_{c0}}{T_{c}})=\Psi(\frac{1}{2}+\frac{\alpha T_{c0}}{2\pi T_{c}})-\Psi(\frac{1}{2})$$
with $\Psi$ being the digamma function, $\alpha = \hbar /2 \tau k_{B}T_{c0}$ the depairing parameter, $\tau$
the elastic scattering time 
and $T_{c0}$ the limit of $T_c$ in the absence of any
scattering. According to fig (\ref{Tcvsrho}) we obtained $T_{c0}$=1.22K from the best least square fit of the data. The critical scattering rate for the
suppression of superconductivity leads to $1/\tau _{cr} = 1.44 cm^{-1}$, $\tau _{cr} = 3.5 ps$ (following the definition of
$\alpha$). Accordingly, $1/\tau$ amounts to $0.13 cm^{-1}$ ($\tau = 40 ps$) in the pristine
\tmc\, sample.

It is now interesting to make the connection with the rate which can be derived from a Drude analysis of the
longitudinal transport using the usual relation $\sigma_{DC}= \omega_{p}^{2}\tau/4\pi$ between transport and plasma frequency data.  Given a 
resistance  ratio of 800 between 300 and 2 K  as currently measured in pure \tmc  in the absence of cracks and a plasma frequency obtained
from the room temperature optical data\cite{Jacobsen83} the Drude mode  
 should have a width at half height of $\approx 2 cm^{-1}$ at low temperature. This width is admittedly much larger than what has been 
derived from the depression of superconductivity by non-magnetic impurities (about 15 times larger). However,
our finding of a very long scattering time in pure \tmc supports the claims made from polarized far-infrared
spectroscopy data
 for a pseudo-gap with a low conductivity in the $2-4 cm^{-1}$  frequency range together with the existence of a zero
frequency peak whose width is less than $0.034 cm^{-1}$ for light //$a$  below 10K \cite{Ng85,Timusk87}. The  DC conductivity at low temperature would in turn
correspond to  the conductivity of this narrow mode at zero frequency.

The suppression of superconductivity at a critical value of the scattering rate corresponds to a critical mean free
path of the order of the microscopic superconducting coherence length ($l\approx \xi_0$). Since the coherence length is anisotropic in such a
low dimensional conductor our data are unable to measure the different components. We have only access to its
average $\xi_{ab}$ in the $a-b^{\prime}$ plane using the Ginzburg-Landau theory and $T_{c}dH_{c2}/dT=\phi _{0}/2\pi
\xi _{ab}^{2}$. Figure (\ref{Hc2vsT}) leads to  $\xi _{ab}=  371\AA$ for \tmc. This value can be compared with
a  determination performed by Murata \textit{et-al} \cite{Murata87} who obtained $\xi_{ab} = 486\AA$.  With fig (\ref{Tcvsrho}), our 
determination of the coherence length gives 
$l\approx 4100\AA$ $(l/\xi_0 = 11$) in \tmc. Hence we can be confident that pristine \tmc \,is located in the clean limit and should be
treated as a clean superconductor.

The data of fig (\ref{Tcvsrho}) and
fig (\ref{Hc2vsT}) cannot be reconciled within the framework of a conventional superconductor. First, the slope $dH_{c2}/dT$ shows a
tendency to decrease together with $T_c$ which is the behaviour expected in a clean type II superconductor when $T_c$ is decreased
but in such a situation $T_c$ itself should not be affected by  non magnetic impurities. Second, in a dirty type II material one can
expect
$dH_{c2}/dT
\alpha \,\rho_{0}$\cite{degennes66} . This is  opposite to the experimental behaviour in fig (\ref{Hc2vsT}).

\vspace{2mm}
In summary, the solid solution \tmx provides an ideal case for the study of the response of organic superconductivity to
non-magnetic disorder. We have shown that on the dilute side with a nominal concentration of $ReO_4$ up to $x=0.06$ the
signature of the anion ordering at 24K provides a proof for the preservation of the original Fermi surface  whereas the
superconducting transition temperature is depressed by a factor about two. This behaviour cannot be reconciled with a model of
conventional superconductors. The gap must show regions of positive and negative signs on the Fermi surface which can be averaged
out by a finite electron lifetime due to elastic scattering. As  these defects are local the scattering momentum of order
$2k_F$ can mix + and - $k_F$ states and therefore the sensitivity to non-magnetic scattering is still unable to tell the difference
between \textit{p} or  \textit{d} orbital symmetry for the superconducting wave function. Furthermore,  the conclusions of our
work corroborate  a previous experimental conclusion showing that the sole existence of a gap anisotropy was unable to explain the
concomitant decreases of both $T_c$ and specific heat anomaly in \tmc \,  when the cooling rate  is increased \cite{Pesty88}.    A noticeable
progress could be achieved paying attention to the spin part of the wave function. In the close vicinity of
$T_c$ orbital limitation for the critical field is expected to prevail and therefore the analysis of the critical fields close to
$T_c$\cite{Gorkov85} does not necessarily imply a triplet pairing. Reliable data in
\tmc\cite{Murata87} are not in contradiction with the picture of singlet pairing  but no data exist below 0.5K, the temperature domain  where
it would be most rewarding to see how $H_{c2}$ compares with the Pauli limit when $H$ is
 perfectly aligned along the
$a$ axis. This new piece of information could select between singlet and triplet pairing hypothesis. The present study has also shown that the
transport in the normal state cannot be explained by a conventional single particle Drude  model. It supports the existence in the
conductivity spectrum of a very narrow zero frequency peak carrying a minor fraction (about
$1\%$) of the total spectral weight as already inferred from the FIR spectroscopy data\cite{Ng85,Timusk87,Schwartz98} and also emphazises the
role of electron correlations in these low dimensional conductors.

\vspace{3mm}
N.Joo acknowledges the french-tunisian cooperation CMCU (project 01/F1303) and thanks S.Haddad, S.Charfi-Kaddour and M.H\'eritier for helpful
discussions.  We also acknowledge a useful discussion with N.Dupuis.


\begin{thebibliography}{10}
\bibitem{noteT2}
We have checked that the inelastic scattering contribution behaving like $T^2$
  in the temperature domain 10K-20K is not affected by the modification of the
  impurity scattering.
\bibitem{note2}
The phase diagram of the R-state of \tmx reported in reference\cite{Tomic83b}
  showing a slight decrease of $T_{AO}$ upon alloying has been obtained with a
  cooling rate five times faster than our smallest cooling rate.
\bibitem{note3}
The determination of the critical temperature in zero field from the
  extrapolation of the critical field could have been performed using any field
  orientation. The $c^*$ orientation has been chosen because it is the axis
  along which it is easier to orient the magnetic field.
\end{thebibliography}

\end{document}